\documentclass[12pt]{article}
\def\R{{\bf R}}
\def\d{\partial}
\def\tr{{\rm tr}}

\begin{document}

\title{Dynamics of $k$-essence}

\author{Alan D. Rendall\\
Max-Planck-Institut f\"ur Gravitationsphysik\\Albert-Einstein-Institut
\\ Am M\"uhlenberg 1\\
14476 Golm, Germany}

\date{}

\maketitle

\begin{abstract}
There are a number of mathematical theorems in the literature on
the dynamics of cosmological models with accelerated expansion driven by a 
positive cosmological constant $\Lambda$ or a nonlinear scalar field with
potential $V$ (quintessence) which do not assume homogeneity and isotropy
from the beginning. The aim of this paper is to generalize these results to 
the case of $k$-essence models which are defined by a 
Lagrangian having a nonlinear dependence on the kinetic energy. In 
particular, Lagrangians are included where late time acceleration is driven 
by the kinetic energy, an effect which is qualitatively different from 
anything seen in quintessence models.  A general criterion for isotropization
is derived and used to strengthen known results in the case of quintessence.
\end{abstract}

\section{Introduction}

In recent years many models have been introduced as possible explanations
of the observed acceleration of the expansion of the universe. The simplest
and best-known of these involve a positive cosmological constant $\Lambda$ or 
a nonlinear scalar field with potential $V$. There is a huge literature on
this subject - see for instance \cite{sahni} for a review. In most papers the 
analysis is restricted to homogeneous and isotropic cosmological models
since it is expected that these will give a good approximate description of
more general spacetimes. For some of the simplest cases with late time
accelerated expansion this expectation has been confirmed by rigorous proofs.
There are a number of papers where mathematical results on the
dynamical properties of models of this kind are obtained in the homogeneous
and inhomogeneous cases. Homogeneous models with $\Lambda>0$ were studied in
\cite{wald} and \cite{lee04a} and those results were extended to scalar 
fields with potentials belonging to various classes in \cite{kitada93}, 
\cite{lee05}, \cite{rendall04a} and \cite{rendall05a}.  In the inhomogeneous
case theorems were proved for vacuum solutions with $\Lambda>0$ in 
\cite{friedrich}, \cite{rendall04b} and \cite{anderson}, the case of an 
exponential potential in \cite{heinzle} and for some inhomogeneous cases 
with matter in \cite{tchapnda03} and \cite{tchapnda05}.
Note also the results of \cite{bieli} on curvature coupled scalar fields.
The purpose of this paper is to extend some of these results to a larger
class of models, the $k$-essence models.

The name $k$-essence denotes a nonlinear scalar field $\phi$ with
Lagrangian $L(\phi,X)$, where $X=-\frac12\nabla_\alpha\phi\nabla^\alpha\phi$,
which is used to describe late-time cosmic acceleration. It was introduced by
Armendariz-Picon, Mukhanov and Steinhardt \cite{armendariz00a} following the 
earlier application of similar models to the very early universe under the 
name $k$-inflation \cite{armendariz99a}. Fields of this kind arise in 
effective field theories coming from string theory when higher order 
corrections in the string tension or number of loops are included. Special 
cases of the $k$-essence Lagrangian are the ordinary minimally coupled 
nonlinear scalar field (associated with the name quintessence) where 
$L(\phi,X)=-V(\phi)+X$ and the tachyon, where $L=-V(\phi)\sqrt{1-2X}$
(see \cite{felder}, \cite{gibbons03a}). Here the potential $V$ is a smooth 
function, assumed in the following to be non-negative. The equation of motion 
is 
\begin{equation}\label{em}
\left(\frac{\d L}{\d X}g^{\alpha\beta}-\frac{\d^2 L}{\d X^2}
\nabla^\alpha\phi\nabla^\beta\phi
\right)\nabla_\alpha\nabla_\beta\phi
+\frac{\d^2 L}{\d\phi\d X}g^{\alpha\beta}\nabla_\alpha\phi\nabla_\beta\phi
=-\frac{\d L}{\d \phi}
\end{equation}
In the spatially homogeneous case this reduces to 
\begin{equation}\label{homem}
\left(\frac{\d L}{\d X}+2X\frac{\d^2 L}{\d X^2}\right)\ddot\phi
+\frac{\d L}{\d X}(3H\dot\phi)+\frac{\d^2 L}{\d\phi\d X}\dot\phi^2
-\frac{\d L}{\d\phi}=0
\end{equation}
Note that in the spatially homogeneous case $X\ge 0$. To ensure the 
solvability of equation (\ref{homem}) for $\ddot\phi$ it should be assumed 
that $\d L/\d X+2X\d^2 L/\d X^2\ne 0$. For a given choice
of $L$ this is a restriction on the allowed values of $\phi$ and $\dot\phi$. 
In general $L$ is not defined on the whole $(\phi,X)$-plane but only on a 
subset of it. As an example, for the tachyon it is necessary that $X<1/2$ in 
order that $L$ be defined. Once this is ensured 
$\d L/\d X+2X\d^2 L/\d X^2>0$ under the sole restriction that the
potential be positive. 

The energy-momentum tensor is given by 
\begin{equation}
T^{\alpha\beta}=\frac{\d L}{\d X}\nabla^\alpha\phi\nabla^\beta\phi
+Lg^{\alpha\beta}
\end{equation}
In the spatially homogeneous case it has two distinct non-vanishing components,
$\rho$ and $p$. In fact the pressure $p$ is just equal to the Lagrangian. 
The energy density $\rho$ is given by $2X\d L/\d X-L$. It 
follows that $\rho+p=2X\d L/\d X$. The inequality $\rho+p\ge 0$ is a not
very restrictive energy condition. In particular it follows from the dominant 
energy condition. If this energy condition is satisfied then it follows that 
$\d L/\d X\ge 0$ in the homogeneous case. In the following it will always be 
assumed that \begin{equation}\label{basics}
\frac{\d L}{\d X}+2X\frac{\d^2 L}{\d X^2}>0, \ \ \ 
\frac{\d L}{\d X}> 0
\end{equation}
As is discussed in Section 2 these conditions follow from the 
hyperbolicity of the equation of motion of $\phi$ and the energy condition
mentioned above. The sign of $\rho+3p=2(X\d L/\d X+L)$ determines whether 
there is accelerated expansion in the homogeneous case. It follows from the 
equation of motion that \begin{equation}\label{dissipate}
\frac{d}{dt}\left(2X\frac{\d L}{\d X}-L\right)=-6HX\frac{\d L}{\d X}
\end{equation}

The paper is organized as follows. Section \ref{inhom} deals with the 
question, for which choices of the 
Lagrangian $L$ the equation of motion of $k$-essence is hyperbolic and 
under what conditions on $L$ various energy conditions are satisfied. In 
Section \ref{rhomin} late time acceleration and isotropization are shown 
for homogeneous spacetimes of Bianchi type I-VIII with $k$-essence defined by 
a Lagrangian chosen from a wide class. The conditions defining this 
class are modelled on those previously used in analysing nonlinear
scalar fields with a potential which have a positive lower bound. In
particular tachyon models whose potential has a positive lower bound
are treated. In these results the spacetime is allowed to contain normal
matter in addition to the $k$-essence. The normal matter is only supposed
to satisfy the dominant and strong energy conditions.
Section \ref{ess} proceeds to consider homogeneous 
spacetimes where late time accelerated expansion is driven by the 
kinetic energy. In these spacetimes the kinetic energy does not go
to zero at late times, in contrast to those studied in Section 
\ref{rhomin}. In Section \ref{isotropize} a general criterion for 
isotropization is obtained and is applied in the special case of quintessence
to generalize the results of \cite{rendall05a}. Appendix 1 contains some 
technical background required for Section \ref{rhomin} while Appendix 2 
contains continuation criteria which are used to obtain global existence 
theorems in Section \ref{rhomin} and \ref{ess}.

\section{The general inhomogeneous case}\label{inhom}

The equations of a $k$-essence field $\phi$ coupled to the Einstein
equations form a well-posed system. This means that if suitable initial
data are prescribed there exists a unique local solution of the 
Einstein-matter equations inducing those initial data. This can be proved 
following the general scheme described in \cite{friedrichr}, section 5.4.
Under the assumptions (\ref{basics}) the equation of motion of $\phi$ is a 
nonlinear wave equation. This follows from the fact that the tensor 
multiplying the second derivatives of $\phi$ in (\ref{em}) is a Lorentz 
metric. The proof of this is given in Appendix 1. By reduction to first 
order this wave equation can be written as a symmetric hyperbolic system. 
The coupling of this system to the Einstein equations is such that the whole 
Einstein-matter system can be written in symmetric hyperbolic form. A local 
existence and uniqueness theorem follows. This argument also applies to the
case where in addition to the coupling to the $k$-essence field the
Einstein equations are coupled to other matter having a well-posed initial
value problem, e.g. a perfect fluid with reasonable equation of state.

One cautionary remark is necessary. For the coupled system initial
data may only be prescribed on a hypersurface which is spacelike with 
respect to both the spacetime metric and the metric in the principal
part of the wave equation for $\phi$. If $\d^2 L/\d X^2\ge 0$ then it is 
enough to assume that the hypersurface is spacelike with respect to
$g_{\alpha\beta}$. Otherwise it is not. In the latter case superluminal
propagation of signals is possible. This is proved in Appendix 1.

Energy conditions can be investigated on a case by case basis as is done
in Appendix 1 for the signature of the metric in the equations of motion. 
The results are as follows. The dominant energy condition is
satisfied iff $\d L/\d X\ge 0$ and $X\d L/\d X-L\ge 0$. For $X>0$
the weak energy condition is equivalent to $\d L/\d X\ge 0$ and
$2X\d L/\d X-L\ge 0$. For $X<0$ the weak energy condition is equivalent 
to $\d L/\d X\ge 0$ and $L\le 0$. The strong energy condition is 
equivalent to $\d L/\d X\ge 0$ and $X\d L/\d X+L\ge 0$ for $X>0$ and
$\d L/\d X\ge 0$ and $-X\d L/\d X+L\ge 0$ for $X<0$.

A potential difficulty with $k$-essence models is that the matter may
form singularities which have nothing to do with the familiar spacetime
singularities of general relativity. (Cf. the discussion in
\cite{felder}.) It would be interesting to know for which
choices of $L$ this can be avoided. A simple place to start is to
look at a $k$-essence field in Minkowski space and consider solutions
evolving from data which are close to zero. A criterion for the corresponding
smooth solution to exist globally in time is the null condition 
\cite{klainerman86a}. This
is an algebraic condition on the nonlinearity. It turns out that it
is satisfied for the $k$-essence field if and only if $\d^k L/\d \phi^k=0$
at $(\phi,X)=(0,0)$ for $k=1,2,3$. In that case there are no singularities 
for small data. In particular this is true if $L$ only depends on $X$.

\section{Solutions with a positive lower bound for $\rho$}\label{rhomin}

Consider a spatially homogeneous solution of Bianchi type I-VIII of the 
Einstein equations with matter. Suppose that the energy-momentum tensor 
is a sum of two parts. One of these $T^M_{\alpha\beta}$ represents ordinary 
matter and satisfies the dominant and strong energy conditions. The other 
represents dark energy and matter quantities derived from it will be given
the subscript $DE$. It is assumed in the following that the dark energy
satisfies the dominant energy condition. Some basic equations are as 
follows:

\begin{eqnarray}
\frac{dH}{dt}&=&-H^2-\frac{4\pi}3(\rho_{DE}+\tr S_{DE})
-\frac13\sigma_{ab}\sigma^{ab}
-\frac{4\pi}3(\rho^M+\tr S^M)\label{evtrk}  \\
H^2&=&\frac{8\pi}3\rho_{DE}+\frac16(\sigma_{ab}\sigma^{ab}-R)
+\frac{8\pi}3\rho^M\label{ham}              \\
\frac{dH}{dt}&=&-4\pi (\rho_{DE}+\frac13\tr S_{DE})
-\frac12\sigma_{ab}\sigma^{ab}+\frac16 R
-4\pi(\rho^M+\frac13\tr S^M)
\label{evtrk2}
\end{eqnarray}
These are the direct generalizations of equations (4), (6) and (7) of
\cite{rendall04a}. Suppose that $H$ is zero at some time $t_1$. Then by 
(\ref{ham}) $\rho_{DE}$ is zero at that time. By (\ref{dissipate}) it 
vanishes at all later times. Since the $k$-essence
field satisfies the dominant energy condition it follows that it makes
no contribution to the energy-momentum tensor. Hence the geometry and
normal matter satisfy the equations in the absence of $k$-essence. By
uniqueness for the full system of equations this must be true at all times
and not just after time $t_1$. It may be said that the $k$-essence field 
is passive and these solutions are physically uninteresting. The results 
of \cite{rendall04a} show that the normal matter vanishes and that spacetime
is flat. For this reason it will be assumed in the following that $H$ is
everywhere positive.

In the next theorem no specific choice of a dark energy model is made. It 
will be specialized to the case of $k$-essence later.

\noindent
{\bf Theorem 1} Consider a solution of the Einstein equations of Bianchi
type I-VIII coupled to a matter model representing dark energy and other
matter satisfying the strong and dominant energy conditions. Let the 
energy-momentum tensor $T^{\alpha\beta}_{DE}$ of the dark energy have
the algebraic form of that of a perfect fluid and satisfy the dominant
energy condition. Suppose that the energy density of dark matter $\rho_{DE}$
has a positive lower bound $\rho_0>0$ and that the solution is initially 
expanding ($H>0$). Then $H\ge H_0$ for some constant $H_0>0$ and if the 
solution exists globally in the future the quantities $\rho^M$, 
$\sigma_{ab}\sigma^{ab}$ and $R$ decay exponentially. 

\noindent
{\bf Proof} The first conclusion follows immediately from the Hamiltonian 
constraint. To prove the remainder of the theorem let $Z=9H^2-24\pi\rho_{DE}$. 
As in the special case of quintessence considered in \cite{rendall04a} the
quantity $Z$ satisfies the inequality $dZ/dt\le -2HZ$. The only properties
of the dark energy model which are used in deriving this are those
included in the hypotheses of the theorem. From the inequality it follows that 
if the solution exists globally in the future the lower bound for $H$ implies
that $Z$ decays exponentially as $t\to\infty$. The Hamiltonian constraint
shows that
\begin{equation}
Z=\frac32\sigma_{ab}\sigma^{ab}-\frac32 R+24\pi\rho^M
\end{equation} 
and this concludes the proof.

\noindent
The interpretation of this theorem is that global solutions isotropize and
that the spatial curvature and the energy density of normal matter
have very little effect on the dynamics at late times. In the case 
of $k$-essence the energy-momentum tensor of a homogeneous model
always has perfect fluid form. To obtain the other conditions it suffices
to supplement  (\ref{basics}) by the inequality $2X\d L/\d X-L\ge C$ for 
a constant $C>0$. The theorem also applies to fluid models for dark
energy which satisfy the dominant energy condition and have a positive
lower bound for the energy density, such as the Chaplygin gas 
\cite{kamenshchik01a}. Note that the argument of the theorem also 
applies to the situation where there is no global positive lower bound for 
$\rho_{DE}$ but for a given solution which exists globally in the future
$\rho_{DE}$ tends to a positive limit as $t\to\infty$.

It follows from equation (\ref{dissipate}) and the lower bound for $H$
that the quantity $X\partial L/\partial X$ is integrable on the positive time 
axis for any global solution.  Now
\begin{equation}
d/dt \left(X\frac{\partial L}{\partial X}\right)=
\dot\phi\left[\ddot\phi\left(\frac{\d L}{\d X}+X\frac{\d^2 L}{\d X^2}\right)
+X\frac{\d^2 L}{\d\phi\d X}\right]
\end{equation}
When the expression for $\ddot\phi$ is substituted into this the 
contribution containing $H$ is non-positive. Discarding it gives an
upper bound for the time derivative of $X\d L/\d X$ in terms of various other 
quantities.

Now a list of relevant assumptions on the Lagrangian will be collected. Given 
a constant $C_1>0$ there exists a constant $C_2>0$ such that:

\begin{enumerate}
\item if $\rho\le C_1$ then $X\le C_2$
\item if $\rho\le C_1$ then $|\d L/\d X+X\d^2 L/\d X^2|\le C_2$
\item if $X\le C_1$ then $|\d L/\d X+2X\d^2 L/\d X^2|\ge C_2^{-1}$
\item if $\rho\le C_1$ then $|\d^2 L/\d\phi\d X|\le C_2$
\item if $\rho\le C_1$ then $|\d L/\d\phi|\le C_2$
\item if $X\le C_1$ then $\d L/\d X\ge C_2^{-1}$
\item there are positive constants $C_3$ and $C_4$ such that $X\le C_3$ 
implies $L(\phi, X)\le -C_4$.
\end{enumerate}

\noindent
{\bf Theorem 2} Consider a solution of the Einstein equations of Bianchi
type I-VIII coupled to a $k$-essence model and other matter satisfying the 
strong and dominant energy conditions. Let the energy-momentum tensor 
$T^{\alpha\beta}_{DE}$ of the $k$-essence satisfy the conditions 
(\ref{basics}) and the dominant energy condition and let there be a positive
lower bound for the energy density. Suppose that the solution is 
initially expanding ($H>0$). Then if the solution exists globally in the 
future and conditions 1.- 7. above are satisfied the expansion is accelerated 
at late times.

\noindent
{\bf Proof} Equation (\ref{dissipate}) shows that $\rho_{DE}$ is bounded
above. Then assumption 1. above implies that $X$ is bounded. Assumptions
2.-5. imply that $X\d L/\d X\to 0$ as $t\to\infty$. By assumption 6.
we can conclude that $X\to 0$ as $t\to\infty$. Finally assumption 7.
implies that the expansion is accelerated at late times. 

Since the assumptions 1.-7. look quite abstract, let us see what they mean 
for quintessence and for the tachyon. Consider first quintessence with a
non-negative potential. Conditions 1.- 4.
obviously hold. Condition 5. holds provided $V'$ is bounded on any interval 
where $V$ is. This assumption played an important role in \cite{rendall04a}. 
Condition 6. is obvious. Condition 7. holds if $V$ has a positive lower bound.
This special case of Theorem 2 was proved in \cite{rendall04a}

Next consider the tachyon. Since $X$ is a priori bounded by one in this case
condition 1. is obvious. In the case of the tachyon if $\rho$ is bounded 
then so is $V$. If in addition $V$ has a positive lower bound then $X$ is 
bounded away from one. Then condition 2. follows. If $V$ has a positive 
lower bound then condition 3. holds. For conditions 4. and 5. it should be 
assumed that $V'$ is bounded whenever $V$ is. With a positive lower bound 
for $V$ condition 6. holds. Condition 7. is automatic for the tachyon.
Thus Theorem 2. implies the following result

\noindent
{\bf Theorem 3} Consider a solution of the Einstein equations of Bianchi
type I-VIII coupled to a tachyon field and other matter satisfying the 
strong and dominant energy conditions. Suppose that the tachyon potential
$V$ has a positive lower bound, that $V'/V$ is bounded and that the solution 
is initially expanding ($H>0$). Then if the solution exists globally in the 
future the expansion is accelerated at late times.

\vskip 10 pt
For a perfect fluid or collisionless matter the global existence assumption
in Theorem 3 holds automatically. For
\begin{equation}
d/dt (\log (1-\dot\phi^2))=(1-\dot\phi^2)(3H\dot\phi^2+V'/V\dot\phi)
\end{equation}
and this prevents $|\dot\phi|$ approaching one in finite time. This 
in turn ensures that $(\phi,\dot\phi)$ remains in a compact subset of the 
domain of definition of $L$. It follows that Lemma 1 and Lemma 2 of 
Appendix 2 can be applied.

If $L$ depends only on $X$ and is globally defined and if it satisfies 
condition 1. then conditions 2.- 6. are automatic. Moreover condition 7. can 
be replaced by the condition $L(0)<0$. In \cite{armendariz99a} the case where 
$L$ only depends on $X$ plays an important role but $L$ does not satisfy the 
positivity conditions (\ref{basics}) everywhere. An alternative interpretation 
is to keep those conditions and say that $L$ is only defined locally. An 
interesting
generalization is given by $L(\phi,X)=L_0(\phi)+L_1(\phi)X+L_2(\phi)X^2$.
This is considered in \cite{armendariz99a} under the condition $L_0=0$ (no 
potential). Theorem 2 applies if $-L_0$ and $L_1$ are bounded below by a 
positive constant.

\section{Solutions with a positive lower bound for the kinetic energy}
\label{ess}

The original aim of $k$-essence models was to produce a scenario where
the accelerated expansion of the universe is driven by the kinetic energy
of a scalar field. This is very different from the usual case of
quintessence where the potential energy is the driving force. In the models 
studied in the previous section it was the latter mechanism which was at
work. The present section is concerned with models closer to the original
motivation for considering $k$-essence. They have the property that they 
have late time accelerated expansion where the kinetic energy does not tend
to zero. 

What can happen is illustrated by the following specific example. Let 
$L=X^2-2X$ on the interval $I=(1,\infty)$. The conditions (\ref{basics})
are satisfied on $I$, as is the dominant energy condition. The equation 
of motion of $\phi$ implies
\begin{equation}
\dot X=-6(3X-1)^{-1}(X-1)XH
\end{equation}
In particular $X$ is decreasing. Also $H\ge (8\pi/3)(3X^2-2X)$. It follows
that if a solution of the Einstein-matter equations exists globally in the 
future then $X\to 1$ as $t\to\infty$. The continuation criteria of Appendix
2 show that when the normal matter is described by a perfect fluid or 
collisionless matter solutions exist globally.  The
mean curvature remains bounded away from zero at late times because
$\rho_{DE}$ does. The inequality $dZ/dt\le 2HZ$ then implies as in the last
section that at late times the influence of the spatial curvature and the
energy density of ordinary matter becomes negligible and the geometry
isotropizes. Since $XL'+L<0$ for $X\in (1,4/3)$ it follows that all solutions 
are accelerated at late times.

This example can easily be generalized. Consider a Lagrangian $L=L(X)$
defined on an interval $(X_0,X_1)$ with $X_0>0$ and $X_1$ finite or infinite.
Suppose further that (\ref{basics}) holds together with the dominant energy 
condition. Finally, suppose that $2XL'-L$ is never zero, that 
$\lim_{X\to X_0}(X-X_0)^{-1}L'$ exists and that $\lim_{X\to X_0}L$ exists and 
is negative. The equation of motion is 
\begin{equation}
\dot X=-6(L'+XL'')^{-1}L'XH
\end{equation}
Thus $X$ is decreasing. From the Hamiltonian constraint it follows that
$H^2\ge (8\pi/3)(2XL'-L)$. Hence under the given conditions if a solution
exists globally in the future then $X\to X_0$ as $t\to\infty$. Provided
$L'+XL''$ is bounded away from zero in a neighbourhood of $X_0$ global
existence theorems follow from the continuation criteria of Appendix 2.
In this case it can be concluded as before that the geometry isotropizes
and that there is late time acceleration. 

In the general case where the Lagrangian also depends on $\phi$ the 
equation of motion implies that
\begin{equation}\label{xdot}
\dot X=-\left(\frac{\d L}{\d X}+2X \frac{\d^2 L}{\d X^2}\right)^{-1}\left[6X
\frac{\d  L}{\d X} H+2\sqrt{2}X^{3/2}\frac{\d^2 L}{\d\phi\d X}
-\sqrt{2}X^{1/2}\frac{\d L}{\d\phi}\right]
\end{equation}
As a global simplifying assumption, suppose that the domain of definition
of $L$ is of the form $J\times I$ where $J$ is an open interval and $I$ is the 
interval $(X_0,\infty)$ with $X_0>0$. Assume as usual 
(\ref{basics}) and the dominant energy condition. In the following a theorem 
is proved which concerns a situation where the dependence of $L$ on $\phi$ 
does not disturb the behaviour seen for a Lagrangian only depending on $X$ 
too much. Looking at (\ref{xdot}) suggests that a smallness assumption should 
be made on $\d L/\d\phi$ and $\d^2 L/\d\phi\d X$. It will be assumed that 
there is a positive constant $\alpha$ such that 
\begin{equation}\label{phideriv}
|\phi\d L/\d\phi|+2|\phi X\d^2 L/\d\phi\d X|\le\alpha (\d L/\d X) 
\end{equation}
This is motivated by the consideration of Lagrangians of the form
$L(\phi,X)=\tilde L(X) \phi^{-2}$ as introduced in \cite{armendariz00a}.
The time derivative of $\phi$ has constant sign and only solutions for 
which $\dot\phi>0$ will be considered. Since a solution which exists
globally in the future satisfies $\dot\phi>\sqrt{2X_0}$ it follows 
that $\phi$ grows at least linearly as $t\to\infty$ so that (\ref{phideriv}) 
implies decay of $(\d L/\d\phi)/(\d L/\d X)$. The following conditions 
are required in the following results. Given a constant $C_1>0$ there exists
a constant $C_2>0$ such that 
\begin{enumerate}
\item if $X\ge X_0+C_1$ then $\d L/\d X\ge C_2^{-1}$
\item $(X-X_0)^{-1}\d L/\d X$ converges to a limiting function as 
$X\to X_0$
\item $L$ converges to a negative limiting function $L_0$ as 
$X\to X_0$
\item if $X\le X_0+C_1$ then $(2X\d L/\d X-L)\ge C_2^{-1}$ 
\item $(\d L/\d X+X\d^2 L/\d X^2)\ge C_2^{-1}$ for all $X$
\item $\d L/\d X\le C_2 X^{1/2}$ for all $X$
\end{enumerate}

\noindent
{\bf Theorem 4} Consider a solution of the Einstein equations of Bianchi
type I-VIII coupled to a $k$-essence model and other matter satisfying the 
strong and dominant energy conditions. Let the energy-momentum tensor 
$T^{\alpha\beta}_{DE}$ of the $k$-essence satisfy the conditions 
(\ref{basics}) and the dominant energy condition. Suppose that the inequality
(\ref{phideriv}) and conditions 1.-4. above are satisfied. If the solution 
is initially expanding and exists globally in the future then the expansion 
is accelerated at late times and the geometry isotropizes.

\noindent
{\bf Proof} Under the assumptions of the theorem the following estimate is
obtained:
\begin{equation}
6X\frac{\d L}{\d X}H+2\sqrt{2}X^{3/2}\frac{\d^2 L}{\d\phi\d X}-\sqrt{2}
X^{1/2}\frac{\d L}{\d\phi}\ge X^{1/2}\frac{\d L}{\d X}(6X^{1/2}H
-\sqrt{2}\phi^{-1}\alpha)
\end{equation}
The function $\phi^{-1}$ tends to zero as $t\to\infty$. Hence the second 
term in the bracket on the right hand side tends to zero. To get 
information about the first term, note that due to (\ref{basics}) the
energy density $\rho_{DE}$ is an increasing function of $X$ and that 
$H\ge\sqrt{8\pi\rho_{DE}/3}$. Together with assumption 4. this implies 
a global positive lower bound for $\rho_{DE}$ and for $H$. Hence 
$\dot X<0$ for $t$ sufficiently large. It follows using assumption 1. that 
$X\to X_0$ as $t\to\infty$. The remaining conclusions of the theorem follow 
using assumptions 2. and 3.

\vskip 10pt
It is of interest to ask when global existence holds so that the theorem 
can be applied. First a lower bound for $X$ will be obtained. At a given time 
either $X<2X_0$ or $X\ge 2X_0$. If the former condition holds let
$t_2$ be the last time before $t_1$ at which $X\ge 2X_0$ if such a time exists 
and let $t_2$ be the initial time otherwise. On this interval $X$ and 
$\phi^{-1}$ are bounded. It follows by condition 5. that
\begin{equation}
|\dot X|\le C\d L/\d X
\end{equation}
for a constant $C$. Since $\d L/\d X$ is $O(X-X_0)$ as $X\to X_0$ it 
can be concluded that $X$ remains bounded away from zero on the interval 
$[t_2,t_1]$ with a bound depending only on the initial data. It follows that 
for any solution $X$ remains bounded away from $X_0$ on any finite time 
interval. Using assumption 6 above and Gronwall's inequality it can
be concluded that $X$ is also bounded on any finite time interval.
It follows that the solution cannot approach the boundary of the domain
of definition of $L$ in finite time. The results of Appendix 2 can be applied 
to obtain global existence statements. 

The conditions 1.-6. do not apply to the case 
$L(\phi,X)=\phi^{-2}\tilde L(X)$ occurring in \cite{armendariz00a}. Next 
some properties of solutions for Lagrangians of this form will be examined. 
The equation of motion takes the form:
\begin{equation}\label{friedman}
(\tilde L'+2X\tilde L'')\dot X+\tilde L'(6XH-4\sqrt{2}\phi^{-1}X^{3/2})
+2\sqrt{2}X^{1/2}\phi^{-1}\tilde L=0
\end{equation}
In this case it is possible for $\dot X$ to become zero, the condition for this
being that $2X\tilde L'-\tilde L=(3/\sqrt{2})\phi X^{1/2}H\tilde L'$. In
a FLRW model without ordinary matter this reduces to
$\tilde L=2X(-6\pi(\tilde L')^2+\tilde L')$.
Any solution of this equation leads to a solution of the Einstein equations
coupled to $k$-essence alone in which $X$ is constant and the scale factor
has power-law behaviour. Solutions of this type were considered in 
section 5 of \cite{armendariz99a} and correspond to the $k$-attractors 
of \cite{armendariz00a}. If $X\tilde L'+\tilde L<0$ the expansion 
is accelerated. Eliminating $\tilde L$ gives the inequality
$\tilde L'>1/4\pi$. The dominant energy condition is equivalent to
$X\tilde L'-L\ge 0$ which gives the inequality $\tilde L'>1/12\pi$.

It is clear that there are many $k$-essence Lagrangians together with
solutions of the type just discussed where $X$ is constant. Note however
that it is not guaranteed that they will exist for any given Lagrangian.
In the case of the tachyon solutions of this kind are known explicitly
\cite{feinstein}. 

By introducing a new time coordinate we can consider (\ref{friedman}) as
an ordinary differential equation for $X$ alone which does not involve 
$\phi$. This makes it very easy to do a local stability analysis of 
the stationary solution in the class of spatially flat Friedmann models
without normal matter. It turns out that the condition for stability is
$\tilde L''(X\tilde L'-\tilde L)>3\pi(\tilde L')^2$. Assuming the 
dominant energy condition this can only be satisfied if $\tilde L''>0$
which means that there is no superluminal propagation in inhomogeneous
models for this Lagrangian. It is more difficult to prove something
about the stability of these solutions within a class of anisotropic
solutions. One reason this is difficult is because the accelerated
expansion is only of power-law type and not faster. In \cite{rendall05a}
isotropization was proved for models whose expansion increases in time
faster than any power of $t$ corresponding to a potential which decays
slower than any exponential. For general potentials which decay at a rate 
comparable to an exponential no result of this kind was obtained. The
only case for which isotropization has been proved is that of an exact
exponential \cite{lee05}. In the next section some partial results on
isotropization are obtained which, in particular, considerably extend
what is known in the case of quintessence.

\section{A criterion for isotropization}\label{isotropize}

Consider a solution of Bianchi type I-VIII of the Einstein equations
coupled to dark energy (satisfying the dominant energy condition)
and ordinary matter (satisfying the dominant and strong energy 
conditions). A computation using the evolution equation for $H$
and the Hamiltonian constraint shows that
\begin{equation}
\frac{dH}{dt}\ge -3H^2\left(1-\frac{4\pi}3\left(\rho_{DE}-\frac13
\tr S_{DE}\right)/H^2\right)
\end{equation}
Suppose that $\liminf_{t\to\infty}(\rho_{DE}-\frac13\tr S_{DE})/H^2>
\frac1{2\pi}$. 
Then it follows that for $t$ sufficiently large $dH/dt\ge -\beta H^2$ for
a constant $\beta<1$. Hence $H(t)\ge\beta^{-1}t^{-1}+O(t^{-2})$. It
follows that for $t$ large $H(t)\ge\gamma t^{-1}$ for any $\gamma$ less
than $\beta^{-1}$. The constant $\gamma$ may be chosen larger than one.
Putting this inequality into the relation $dZ/dt\le -2HZ$ shows that
$Z=O(t^{-2\gamma})$ and that $Z/H^2\to 0$ as $t\to\infty$. Thus 
a criterion for isotropization has been obtained.

What does the criterion look like in the case of quintessence? There
\begin{equation}
\left(\rho_{DE}-\frac13\tr S_{DE}\right)/H^2=\frac{2V}{H^2}
\end{equation}
so that we get the inequality $\liminf (8\pi V/3H^2)>2/3$. It will now 
be shown that if $\alpha=\limsup_{\phi\to\infty}(-V'/V)<4\sqrt{\pi/3}$ then 
the criterion is satisfied. The starting point of the proof is provided by 
equation (11) and the inequality (13) of \cite{rendall05a}. There is a 
number $\phi_1$ such that $-V'/V\le \alpha/2+\sqrt{4\pi/3}$ for all 
$\phi\ge\phi_1$.
Now distinguish between the cases $\dot\phi/H\ge 1/\sqrt{12\pi}$ and
$\dot\phi/H\le 1/\sqrt{12\pi}$. In the first case equation (11) of
\cite{rendall05a} gives an upper bound for $d/d\phi(3H^2/8\pi V)$
while in the second case equation (11) and the inequality (13) of that
reference imply the bound (14) for $d/d\phi(3H^2/8\pi V)$. Combining
these two bounds shows that the following inequality holds in general:
\begin{equation}
\frac{d}{d\phi}\left(\frac{3H^2}{8\pi V}\right)\le -\min\left\{\left[
\sqrt{48\pi}\left(1-\frac{8\pi V}{3H^2}\right)+\frac{V'}{V}\right],
\sqrt{\frac{4\pi}{3}}-\frac{\alpha}{2}\right\}
\left(\frac{3H^2}{8\pi V}\right)
\end{equation}
Suppose that 
\begin{equation}
\frac{3H^2}{8\pi V}\ge\left(1-\frac{\alpha}{\sqrt{48\pi}}\right)^{-1}+C_1
\end{equation}
on some interval for a positive constant $C_1$. Then
\begin{equation}
1-\frac{8\pi V}{3H^2}\ge \left[C_1\left(1-\frac{\alpha}{\sqrt{48\pi}}\right)
+\frac{\alpha}{\sqrt{48\pi}}\right]\left[1+C_1\left(1-
\frac{\alpha}{\sqrt{48\pi}}\right)\right]^{-1}
\end{equation}
The expression on the right hand side of this inequality is strictly
greater than $\alpha/\sqrt{48\pi}$. Increase $\phi_1$ if necessary so that 
\begin{equation}
\left|\frac{V'}{V}\right|\le\frac12\sqrt{48\pi}
\left[C_1\left(1-\frac{\alpha}{\sqrt{48\pi}}\right)
+\frac{\alpha}{\sqrt{48\pi}}\right]\left[1+C_1\left(1-
\frac{\alpha}{\sqrt{48\pi}}\right)\right]^{-1}
\end{equation}
for $\phi>\phi_1$. This is possible due to the restriction on $\alpha$
which has been assumed. It follows that the criterion for isotropization
holds.

The result just proved extends the proof of isotropization and decay of 
spatial curvature and energy density of ordinary matter significantly
in comparison with what is proved in \cite{rendall05a}. It applies for 
example to the potentials $V(\phi)=V_0(\log\phi)^p\phi^n\exp (-\lambda\phi)$,
the special case of equation (18) in \cite{rendall05a} with $m=1$
provided $\lambda<4\sqrt{\pi/3}$ and to a number of other potentials
listed in \cite{rendall05a}.

\section{Conclusions and outlook}

In this paper results are obtained on the dynamics of solutions of the
Einstein equations with $k$-essence and normal matter. They include
anisotropic spacetimes of Bianchi type I-VIII and not only FLRW models.
In this setting late time isotropization is proved. They allow the normal 
matter to be anything which satisfies the strong and dominant energy 
conditions. Thus the results are not confined to fluids or multifluids
and apply, for instance, to collisionless matter. All the arguments are
rigorous and not dependent on heuristics.

Before coming to the global results for homogeneous solutions, basic 
information is presented concerning the initial value problem for the 
Einstein-matter equations with $k$-essence in general, i.e. without symmetry 
assumptions. A link is made with a concept in the theory of nonlinear 
hyperbolic equations, the null condition, and it would be desirable to
establish further connections in the future. The paper \cite{felder}
could be a good starting point.

It is shown that in general if $\rho_{DE}$ satisfies a suitable lower
bound homogeneous solutions of Bianchi type I-VIII which exist globally
in the future become homogeneous with dimensionless measures of the
spatial curvature and density of normal matter tending to zero
exponentially. This works if a global lower bound is available for
the given matter model or if it is available for a particular solution.
It is shown further in Section 3 that if certain bounds are assumed on the 
Lagrangian it is possible to prove late-time accelerated expansion. In
this class of models $X\to 0$ as $t\to\infty$. They very much resemble
quintessence models with a lower bound for the potential. It would be
interesting to prove a similar result for a class generalizing 
quintessence models with a potentials tending to zero at infinity
as analysed in \cite{rendall05a}.

In Section 4 results are obtained on late time behaviour of solutions
whose kinetic energy does not tend to zero as $t\to\infty$. They
achieve the aim of proving late time accelerated expansion and 
isotropization for models where these phenomena occur. The questions
of behaviour at intermediate times and late time behaviour without
acceleration are not addressed since they are not amenable to the 
techniques used here. If the expansion is not accelerated at late 
times then the behaviour is strongly dependent on Bianchi type and
a unified treatment is difficult or impossible. As far as the
behaviour at intermediate times is concerned it is difficult to
even formulate interesting and precisely defined statements 
susceptible to rigorous proof. This is connected to the fact that
the interesting statements often involve essentially quantitative
elements while rigorous results tend to be qualitative in nature.
These issues deserve a separate discussion.   

The first case considered in Section 4 is that where $L$ is independent
of $\phi$. A more general theorem is then proved which shows that certain
kinds of dependence on $\phi$ do not change the qualitative behaviour in
comparison to the case where $L$ only depends on $X$. Late time acceleration
and isotropization are proved. Next the case $L(\phi,X)=\phi^{-2}\tilde L(X)$
is considered. This allows solutions with $\dot X=0$, the $k$-attractors.
The stability of these solutions within the class of isotropic spatially
flat models without ordinary matter is determined. Isotropization is not 
proved. The fact that the expansion of the candidate attractor is only
of power-law type makes this difficult, as it does for quintessence
models with this type of asymptotics. To make further progress a
criterion for isotropization in dark energy models is developed in
Section 5. It is shown that this criterion leads to a significant
improvement in the results which can be obtained in the case of
quintessence. Similar conclusions have not yet been obtained for more 
general $k$-essence models. 

This paper contains new results on the mathematical properties of
solutions of the Einstein equations coupled to dark energy and other
matter and concentrates on establishing a basis for the mathematical
study of $k$-essence models. The $k$-essence Lagrangians give rise
to a considerable variety of behaviour and the task of obtaining
an overview of the possibilities is a challenge for the future. 

\vskip 10pt\noindent
{\bf Appendix 1. Analysis of characteristics}

\noindent
Results similar to those of this appendix were obtained in 
\cite{armendariz05a}. The treatment here is intended to make the proofs
more transparent by avoiding the use of series expansions.

Consider the expression 
$\frac{\d L}{\d X}g^{\alpha\beta}-\frac{\d^2 L}{\d X^2}
\nabla^\alpha\phi\nabla^\beta\phi$
which occurs in the principal part of the $k$-essence equations of motion.
Evidently this is degenerate if $\d L/\d X=0$. We now assume that
$\d L/\d X\ne 0$ and define 
\begin{equation}
\tilde h^{\alpha\beta}=(\d L/\d X)^{-1}(\frac{\d L}{\d X}g^{\alpha\beta}
-\frac{\d^2 L}{\d X^2} \nabla^\alpha\phi\nabla^\beta\phi)
=g^{\alpha\beta}-K\nabla^\alpha\phi\nabla^\beta\phi
\end{equation}
where $K=\frac{\d^2 L/\d X^2}{\d L/\d X}$.
If $1-K\nabla_\sigma\phi\nabla^\sigma\phi$ is non-zero let
$h_{\alpha\beta}=g_{\alpha\beta}+\frac{K\nabla_\alpha\phi\nabla_\beta\phi}
{1-K\nabla_\sigma\phi\nabla^\sigma\phi}$. Then by a direct computation 
$h_{\alpha\beta}$ is the inverse of $\tilde h^{\alpha\beta}$. The
condition $1-K\nabla_\sigma\phi\nabla^\sigma\phi=0$ is equivalent to
$\frac{\d^2 L}{\d X^2}\nabla_\sigma\phi\nabla^\sigma\phi=\frac{\d L}{\d X}$.
If $\frac{\d^2 L}{\d X^2}=0$ then this never happens. Otherwise the assumption
that it does not happen puts a restriction on 
$\nabla_\sigma\phi\nabla^\sigma\phi$.

Next the signature of the metric $\tilde h^{\alpha\beta}$ will be 
investigated. Suppose first that the gradient of $\phi$ is timelike.
Then $\nabla^\alpha\phi=AT^\alpha$ for a unit timelike vector $T^\alpha$.
Then $\tilde h^{\alpha\beta}T_\alpha T_\beta=-1-KA^2$ while
$\tilde h^{\alpha\beta}S_\alpha S_\beta=1$ for any unit spacelike vector
$S^\alpha$ orthogonal to $T^\alpha$. Moreover 
$\tilde h^{\alpha\beta}S_\alpha T_\beta=0$. Thus the signature of
$\tilde h$ is Lorentzian for $KA^2>-1$. For $KA^2<-1$ it is positive
definite. The equation becomes elliptic. If the gradient of $\phi$ is 
spacelike then it follows that $\nabla^\alpha\phi=AS^\alpha$ for a unit
spacelike vector $S^\alpha$. A calculation similar to the one just
done shows that the signature is Lorentzian for $KA^2<1$ and 
$(-,-,+,+)$ for $KA^2>1$. If $\nabla^\alpha\phi$ is null the signature is 
always Lorentzian. Hence in all cases an equivalent condition to the
metric being Lorentzian is $K\nabla_\alpha\phi\nabla^\alpha\phi<1$.
This can be re-expressed as $2X\frac{\d^2 L}{\d X^2}+\frac{\d L}{\d X}>0$
when $\d L/\d X>0$.

Let $v^\alpha$ be a vector which is null with respect to the metric 
$h_{\alpha\beta}$. Then $g_{\alpha\beta}v^\alpha v^\beta=-K(\nabla_\alpha\phi
v^\alpha)^2/(1-K\nabla_\alpha\phi\nabla^\alpha\phi)^2$. It follows that
if $K$ is positive the null cone of $h_{\alpha\beta}$ is inside (or on) 
that of $g_{\alpha\beta}$ while if $K$ is negative it is outside (or on)
that of $g_{\alpha\beta}$.

\vskip 10pt\noindent
{\bf Appendix 2. Continuation criteria}

\noindent
In this appendix basic local existence theorems and continuation criteria
will be proved for homogeneous spacetimes containing $k$-essence and normal
matter. The choices of normal matter discussed are a large class of perfect
fluids with positive pressure, a mixture of several non-interacting perfect 
fluids and collisionless matter described by the Vlasov equation. Let $G$ be 
the open subset of $\R^2$ where the Lagrangian $L$ of the $k$-essence field 
is defined. Assume that the equation of state $p=f(\rho)$ of any perfect 
fluid occurring is such that $f:[0,\infty)\to [0,\infty)$ is continuous with
$f(0)=0$ and that for $\rho>0$ the function $f$ is $C^1$ with 
$0\le f'(\rho)\le 1$

\noindent
{\bf Lemma 1} Consider a solution of Bianchi type I-VIII of the Einstein 
equations coupled to a perfect fluid and a $k$-essence field $\phi$ on a time 
interval $[t_0,t_1)$ with $t_1<\infty$. Suppose that the $k$-essence field 
satisfies the dominant energy condition. If $H(t)$ is bounded and if 
$(\phi,\dot\phi^2/2)$ remains in a compact subset of $G$ then the solution 
can be extended to a longer time interval. 

\noindent
{\bf Proof} The proof follows the discussion in section 4 of \cite{rendall95a}.
The equations of motion of the fluid and the $k$-essence field can be written 
in the form 
\begin{equation}
A(z,g_{ij},k_{ij})dz/dt=F(z,g_{ij},k_{ij})
\end{equation}
where $z=(\rho, u^i, \phi, \dot\phi)$ and tensors are expressed in components 
with respect to a left invariant frame of the Lie algebra defining a given 
Bianchi type. The matrix $A$ is invertible. The standard local existence 
theorem for ordinary differential equations \cite{hartman} applies to this 
system. It gives local existence of solutions and shows that the solution 
can be extended to a longer time interval provided $(z,g_{ij},k_{ij})$ remains 
within a compact subset of the domain of definition of the coefficients. It 
will be shown that this follows from the assumptions of the Lemma. A first 
step is to show that under the given assumptions $g_{ij}$, $(\det g)^{-1}$ and 
$k_{ij}$ remain bounded. The proof adapts an argument used in the proof
of Lemma 2.1 of \cite{rendall94a}. There the boundedness of the integral
of $k_{ij}k^{ij}$ was deduced from certain energy conditions. In fact the
dominant energy condition alone is enough. Equation (\ref{evtrk2}) can
be used to bound the integral of $\sigma_{ij}\sigma^{ij}$ where $\sigma_{ij}$
is the tracefree part of $k_{ij}$. Together with the assumed boundedness of
$H$ this gives the desired conclusion. It remains to show that $\rho$, 
$\rho^{-1}$ and $u^i$ are bounded.. This can be done just as in 
\cite{rendall95a}.

\vskip 10pt
As in \cite{rendall95a}, this result can immediately be extended to the case
of several non-interacting fluids, e.g. a mixture of dust and radiation. It 
can also be adapted to the case of collisionless matter. For this purpose
the existence theorem of section 2 of \cite{rendall94a} has to be extended
but this is straightforward. The quantities $\phi$ and $\dot\phi$ have to be 
added to those for which an iteration is carried out. The result is:

\noindent
{\bf Lemma 2} Consider a solution of Bianchi type I-VIII of the Einstein 
equations coupled to a collisionless gas described by the Vlasov equation 
and a $k$-essence field $\phi$ on a time interval $[t_0,t_1)$ with 
$t_1<\infty$. Suppose that the $k$-essence field satisfies the dominant 
energy condition. If $H(t)$ is bounded and if $(\phi,\dot\phi^2/2)$ remains 
in a compact subset of $G$ then the solution can be extended to a longer 
time interval.

\end{document}